\documentstyle[12pt,epsfig,twoside]{article}
\newcommand{\beq}{\begin{equation}}
\newcommand{\eeq}{\end{equation}}
\newcommand{\beqn}{\begin{eqnarray}}
\newcommand{\eeqn}{\end{eqnarray}}
\newcommand{\vs}{\\[0.3cm]\indent}
\newcommand{\intl}{\int\limits}
\newcommand{\mc}{\multicolumn}

\def\tauto{$\tau^{-\!}\rightarrow\,$}
\def\nut{$\,\nu_\tau$}
\def\piz{$ \pi^0 $}
\def\pipiz{$ \pi^-\pi^0 $}
\def\pitpiz{$ \pi^-3\pi^0 $}
\def\tpipiz{$ 2\pi^-\pi^+\pi^0 $}

\def\Ks{K$^0_{\mathrm S}$}
\def\Kl{K$^0_{\mathrm L}$}

\def\aqed{$\alpha(s)$}
\def\aqedZ{$\alpha(M_{\mathrm Z}^2)$}
\def\daqed{$\Delta\alpha(s)$}
\def\daqedZ{$\Delta\alpha(M_{\mathrm Z}^2)$}

\def\daqedhZ{$\Delta\alpha_{\mathrm had}^{(5)}(M_{\mathrm Z}^2)$}
\def\amuhad{$a_\mu^{\mathrm had}$}

\def\sf{spectral function}
\def\sfs{spectral functions}

\def\ee{$e^+e^-$}

\def\pc{$\%$}
\def\rs{\raisebox{1.5ex}[-1.5ex]}
\def\via{via}
\def\pms{$\,\pm\,$}

\def\GeVM{~GeV$/c^2$}
\def\GeVM2{~GeV$^2/c^4$}

\def\MeVM{~MeV$/c^2$}

\begin{document}

\begin{titlepage}
\setcounter{page}{1}

\begin{flushright} 
{\bf LAL 97-02}\\
{February 1997}
\end{flushright} 

\vspace{1cm}

\begin{center} 
\begin{Large}
{\bf Improved Determination of the Hadronic Contribution to the Muon 
     (\boldmath$g-2$) and to \boldmath$\alpha(M_{\mathrm Z}^2)$ \\
     Using new Data from Hadronic \boldmath$\tau$ Decays} \\
\end{Large}
\vspace{1.6cm}
\begin{large}
Ricard Alemany$^{\,\mathrm a,}$\footnote{E-mail: alemany@alws.cern.ch},
Michel Davier$^{\,\mathrm b,}$\footnote{E-mail: davier@frcpn11.in2p3.fr}
and Andreas H\"ocker$^{\,\mathrm b,}$\footnote{E-mail: hoecker@lalcls.in2p3.fr} \\
\end{large}
\vspace{0.5cm}
{\small \em $^{\mathrm a}$CERN, Switzerland}\\
\vspace{0.1cm}
{\small \em $^{\mathrm b}$Laboratoire de l'Acc\'el\'erateur Lin\'eaire,\\
IN2P3-CNRS et Universit\'e de Paris-Sud, F-91405 Orsay, France}\\
\vspace{3.cm}

{\small{\bf Abstract}}
\end{center}
{\small
\vspace{-0.2cm}
We have reevaluated the hadronic contribution to the anomalous magnetic 
moment of the muon $(g-2)$ and to the running of the QED f\/ine structure
constant \aqed\ at $s=M_{\mathrm Z}^2$. We incorporated new data from hadronic 
$\tau$ decays, recently published by the ALEPH Collaboration. In addition,
compared to previous analyses, we use more extensive \ee\ annihilation data 
sets. The integration over the total hadronic cross section is performed 
using experimental data up to 40~GeV and results from perturbative QCD above 
40~GeV. The improvement from $\tau$ data concerns mainly the pion form factor, 
where the uncertainty in the corresponding integral could be reduced by more
than a factor of two. We obtain for the lowest order hadronic vacuum 
polarization graph \amuhad\,=\,(695.0$\,\pm\,$15.0)$\times10^{-10}$ and
\daqedhZ\,=\,(280.9$\,\pm\,$6.3)$\times10^{-4}$ using \ee\ data only. The
corresponding results for combined \ee\ and $\tau$ data are 
\amuhad\,=\,(701.1$\,\pm\,$9.4)$\times10^{-10}$ and 
\daqedhZ\,=\,(281.7$\,\pm\,$6.2) 
$\times10^{-4}$, where the latter is calculated using the contribution from
the five lightest quarks.
\noindent
}
\vspace{25mm}

\end{titlepage}

\newpage\thispagestyle{empty}.\newpage
\setcounter{page}{1}
%
%
\section*{Introduction}
\label{sec_introduction}

The anomalous magnetic moment $a_\mu=(g-2)/2$ of the muon is experimentally
and theoretically known to very high accuracy. In addition, the contribution 
of heavier objects to $a_\mu$ relative to the anomalous moment of the electron 
scales as $(m_\mu/m_e)^2\sim4\times10^{4}$. These properties allow an 
extremely sensitive test of the validity of QED and of additional contributions 
from strong and electroweak interactions. The present value from the combined 
$\mu^+$ and $\mu^-$ measurements~\cite{bailey},
\beq
    a_\mu \:=\: (11\,659\,230 \pm 85)\times10^{-10}~,
\eeq
should to be improved to a precision of at least
$4\times10^{-10}$ by a forthcoming Brookhaven experiment (BNL-E821)~\cite{bnl}.

It is convenient to separate the prediction from the Standard Model (SM) 
$a_\mu^{\mathrm SM}$ into its dif\/ferent contributions
\beq
    a_\mu^{\mathrm SM} \:=\: a_\mu^{\mathrm QED} + a_\mu^{\mathrm had} +
                             a_\mu^{\mathrm weak}~,
\eeq
where $a_\mu^{\mathrm QED}=(11\,658\,470.6\,\pm\,0.2)\times10^{-10}$ is 
the pure electromagnetic contribution (see~\cite{krause1} and references 
therein), \amuhad\ is the contribution from hadronic vacuum polarization,
and $a_\mu^{\mathrm weak}=(15.1\,\pm\,0.4)\times10^{-10}
$~\cite{krause1,kuraev,weinberg} accounts for corrections due to the 
exchange of the weak interacting bosons up to two loops. Using the recent 
analysis of S.~Eidelman and F.~Jegerlehner~\cite{eidelman} that found
\amuhad$\,=\,(702.4\,\pm\,15.3)\times10^{-10}$, and applying fourth order 
corrections due to the exchange of additional photons and electron or quark 
loops (summarized by T.~Kinoshita {\it et al.}~\cite{kinoshita} to 
$(-4.1\,\pm\,0.7)\times10^{-10}$) one f\/inds
\beq
    a_\mu^{\mathrm SM} = (11\,659\,184 \,\pm\, 16)\times10^{-10}~.
\eeq
Comparing the errors of the respective contributions to $a_\mu^{\mathrm SM}$ 
reveals that its total uncertainty is clearly dominated by
the leading order vacuum polarization correction \amuhad, originating from
a quark-loop insertion into the muon vertex correction diagram as shown
in F\/igure~\ref{fig_amu}. 
\vs
In this letter, we present a new evaluation of the hadronic vacuum 
polarization contribution to $a_\mu$ and also to the running of the 
QED f\/ine structure constant $\alpha(s)$ from low energy to the mass 
of the Z boson. In addition to using the complete and in comparison 
with previous analyses slightly enlarged experimental information 
on \ee\ annihilation data, we incorporate new data from hadronic 
$\tau$ decays~\cite{Aleph} which provide a more precise description
of the hadronic contributions at energies less than 1.5~GeV. We bring 
attention to the straightforward and statistically well-def\/ined 
averaging procedure and error propagation used in this paper, which 
takes into account full systematic correlations between the cross 
section measurements. We also stress our careful treatment of unmeasured 
f\/inal states which are bound \via\ isospin constraints.

%
%
\section{Hadronic Vacuum Polarization in \boldmath$\gamma$ and W propagators}
\label{sec_cvc}

As QCD is a non-Abelian theory with massless gauge bosons, its perturbative
expansion at low energies is not well-behaved and non-perturbative ef\/fects
lead to currently unpredictable long distance resonance phenomena in
quark interactions. Fortunately, cross sections measured in \ee\ annihilation
and \sfs\ from $\tau$ decays provide an experimental access to the hadronic 
vacuum polarization: from unitarity, the hadronic cross section of \ee\ 
annihilation is related to the absorptive part of the vacuum polarization 
correlator \via\ the optical theorem. 
\vs
Similarly, hadronic \sfs\ from $\tau$ decays are directly related to the 
isovector vacuum polarization currents when isospin invariance (CVC) and 
unitarity hold. For this purpose we have to worry whether the breakdown 
of CVC due to quark mass ef\/fects ($m_u \neq m_d$ generating 
$\partial_{\mu}J^\mu\sim(m_u - m_d)$ for a charge-changing hadronic 
current $J^\mu$ between $u$ and $d$ quarks) or unknown isospin-violating 
electromagnetic decays has non-negligible contributions within the 
present accuracy. Recent estimates of CVC predictions of $\tau$ branching 
ratios into vector f\/inal states~\cite{eidelmantau} however show good 
agreement within about 5\pc\ experimental accuracy over the full range 
of exclusive vector hadronic f\/inal states. Expected deviations from CVC 
due to so-called {\it second class currents} as, {\it e.g.}, the decay 
$\tau^-\rightarrow\pi^-\eta$\nut\ where the corresponding \ee\ f\/inal state
\piz$\eta$ (C=+1) is strictly forbidden, have estimated branching 
fractions of order of $(m_u-m_d)^2\simeq10^{-5}$~\cite{etapi}, while the 
experimental upper limit amounts to 
B($\tau\rightarrow\pi^-\eta$\nut)\,$\,<1.4\times10^{-4}$~\cite{PDG} with 95\pc\ 
CL. Another classical test is the pion $\beta$-decay, yielding a sensitivity 
to CVC violation of 3\pc. The CVC hypothesis relates the isovector, vector 
matrix element of the decay $\pi^-\rightarrow\pi^0e^-\bar{\nu}_e$ to the 
electromagnetic form factor of the pion. No deviation between the CVC 
branching ratio and the experimental result has been found~\cite{cvc1,PDG}. 
\vs
An estimate of a possible CVC violation can be obtained in the $\pi\pi$
f\/inal state which is dominated by the $\rho(770)$ resonance. $SU(2)$ 
symmetry breaking can occur in the $\rho^\pm$--$\rho^0$ masses and 
widths caused by electromagnetic interactions. Hadronic contributions to 
the $\rho^\pm$--$\rho^0$ width dif\/ference are expected to be much 
smaller since they are proportional to $(m_u-m_d)^2$. The dif\/ferent 
electromagnetic contributions to the width are listed in 
Table~\ref{tab_cvcbreak}:
radiative transitions introduce a negligible ef\/fect, while the
dominant contribution comes from the $\pi^\pm$--$\pi^0$ mass dif\/ference.
A recent theoretical analysis~\cite{bijnens} indicates that the 
$\rho^\pm$ and $\rho^0$ masses are in fact equal within 0.5\MeVM:
$m_{\rho^\pm}-m_{\rho^0}=-(0.15\pm0.55)$\MeVM. This prediction has
been verif\/ied in the ALEPH analysis of the $\tau$ vector 
\sfs~\cite{Aleph} with the result: 
$m_{\rho^\pm}-m_{\rho^0}=-(0.0\pm1.0)$\MeVM. Since the $\pi^-$ and
$\pi^0$ mass dif\/ference is known experimentally~\cite{PDG} to
be $m_{\pi^\pm}-m_{\pi^0}=(4.5936\pm0.0005)$\MeVM\ and understood
theoretically~\cite{gasser_leut} to be almost completely from 
electromagnetic origin, it is expected that the total $\rho^\pm$
and $\rho^0$ widths should be dif\/ferent even in the limit where
hadronic interactions are $SU(2)$-invariant (this includes the 
chiral limit). An additional point concerns contributions from 
photon bremsstrahlung. While the infrared divergences 
in $\rho\rightarrow\pi\pi\gamma$ decays vanishes when including the 
vertex correction graphs, f\/inite terms are expected to contribute 
dif\/ferently to the widths of the charged and the neutral $\rho$. 
The corresponding bremsstrahlung graphs have been calculated in 
Ref.~\cite{singer}. The width contributions from f\/inite terms to both 
the $\rho^\pm$ and the $\rho^0$ turn out to be negative. The estimate 
of the width dif\/ference given in Tab.~\ref{tab_cvcbreak} assumes 
f\/inite contributions to the widths from loop corrections to be small.
\vs
The total expected $SU(2)$ violation in the $\rho$
width is f\/inally computed from the above considerations to be
\beq
\label{eq_gamdif}
   \frac{\Gamma_{\rho^\pm}-\Gamma_{\rho^0}}{\Gamma_{\rho}} \;=\;
      (2.8 \pm 3.9)\times 10^{-3}~,
\eeq
where the error comes essentially from the estimate of the
$\rho^\pm$--$\rho^0$ mass dif\/ference and of the $\pi\pi\gamma$
contribution. This ef\/fect introduces corrections for 
$a_\mu^{\rm had}$ and the running of $\alpha(s)$ when including
$\tau$ data (see Section~\ref{sec_results}).
\begin{table}[t]
\begin{center}
\begin{tabular}{ |c|c|c| } \hline && \\
  \rs{F\/inal states}
& \rs{$\frac{\Gamma_{\rho^\pm}-\Gamma_{\rho^0}}{\Gamma_{\rho}}$  ($\times10^3$)}
                                                          & \rs{Ref.}      \\ 
\hline\hline
$\pi\omega\rightarrow\pi\pi^0\gamma$  &    0.32           & \cite{singer}  \\
$\pi\gamma$                           & $-$0.34\pms0.21   & \cite{PDG}     \\
$\eta\,\gamma$                        & $-$0.38\pms0.07   & \cite{PDG}     \\
$\ell^+\ell^-$                        & $-$0.091\pms0.004 & \cite{PDG}     \\
$m_{\pi^\pm}-m_{\pi^0}$, $m_{\rho^\pm}-m_{\rho^0}$                
                                      &    6.3\pms2.5     & \cite{PDG,bijnens} \\
$\pi\pi\gamma$                        & $-$3\pms3         & \cite{singer}  \\
\hline \hline
Sum                                   & \mc{1}{c}{2.8\pms3.9} & \mc{1}{c|}{}\\
\hline
\end{tabular}
\end{center}
\caption{\label{tab_cvcbreak}\it
         Expected CVC violation from electromagnetic interactions
         in $\rho^\pm$--$\rho^0$ decays.}
\end{table} 
\vs
From a more qualitative point of view, one should keep in mind that in this 
analysis the CVC hypothesis is applied at a very low energy scale where 
the absorptive parts of the matrix elements are largely dominated by 
non-perturbative QCD which are expected to factorize from their respective 
W or $\gamma$ excitation. 
\vs
However, electroweak radiative corrections must be taken into account. 
Their dominant contribution comes from the short distance correction
to the ef\/fective four-fermion coupling $\tau^-\rightarrow (d\bar{u})^-$\nut\
yielding the fractional logarithmic term 
$(\alpha(M_\tau)/2\pi)({4\,\mathrm{ln}}(M_{\mathrm Z}/M_\tau)$ 
$+5/6)$~\cite{braaten} which vanishes in leptonic $\tau$ decays. The short 
distance correction can be absorbed into an overall multiplicative electroweak 
correction $S_{\mathrm{EW}}$ introduced in the def\/inition of the \sfs\ 
used in Ref.~\cite{Aleph}. The resummation of higher-order electroweak 
logarithms using the renormalization group yields~\cite{bnp,marciano}
\beq\label{eq_sew}
    S_{\mathrm{EW}} \;=\; 
 \left(\frac{\alpha(m_b)}{\alpha(M_\tau)}\right)^{\!\!\frac{9}{19}}\times
 \left(\frac{\alpha(M_{\mathrm W})}{\alpha(m_b)}\right)^{\!\!\frac{9}{20}}\times
 \left(\frac{\alpha(M_{\mathrm Z})}{\alpha(M_{\mathrm W})}\right)^{\!\!\frac{36}{17}}
   \,=\, 1.0194~,
\eeq
while remaining perturbative electroweak corrections are of order 
$\alpha^n(M_\tau)\,{\mathrm{ln}}^n(M_{\mathrm Z}/M_\tau)$ which is safe to 
ignore. The subleading non-logarithmic short distance correction, calculated
to order $O(\alpha)$ at quark level: 
$5\alpha(M_\tau)/12\pi\simeq0.0010$~\cite{braaten} turns out to be small.
Additional long-distance corrections are expected to be f\/inal state dependent.
They have only been computed for the \tauto$\pi^-$\nut\ decay leading to
a total radiative correction of 2.03\pc~\cite{decker} which is dominated by 
the leading logarithm from the short distance contribution.
The evaluation of~(\ref{eq_sew}) neglects radiative corrections from 
additional gluon exchange between the f\/inal state quarks. 
\vs
To be safe~\cite{private}, the uncertainty of $S_{\mathrm{EW}}$ 
in~(\ref{eq_sew}) is estimated to be $\Delta S_{\mathrm{EW}}\,=\,0.0040$, 
which is taken into account in the CVC cross section prediction from $\tau$ 
decays. 

%
%
\subsection*{Muon Magnetic Anomaly}

By virtue of the analyticity of the vacuum polarization correlator, the 
contribution of the hadronic vacuum polarization to $a_\mu$ can be 
calculated \via\ the dispersion integral~\cite{rafael}
\beq\label{eq_integral1}
    a_\mu^{\mathrm had} \:=\: 
           \frac{1}{4\pi^3}
           \intl_{4m_\pi^2}^\infty ds\,\sigma_{\mathrm had}(s)\,K(s)~.
\eeq
Here $\sigma_{\mathrm had}(s)$ is the total \ee$\rightarrow\,$hadrons 
cross section as a function of the c.m. energy-squared $s$, and
\begin{figure}[t]
\epsfxsize4.5cm
\centerline{\epsffile{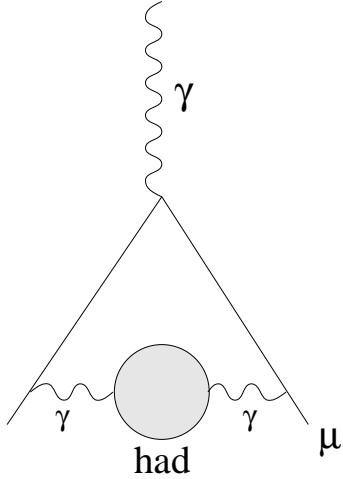}}
\caption{\it Leading order hadronic vacuum polarization contribution
         to $a_\mu$.}
\label{fig_amu}
\end{figure}
$K(s)$ denotes the QED kernel
\beq
      K(s) \:=\: x^2\left(2-\frac{x^2}{2}\right) \,+\,
                 (1+x)^2\left(1+\frac{1}{x^2}\right)
                      \left({\mathrm ln}(1+x)-x+\frac{x^2}{2}\right) \,+\,
                 \frac{(1+x)}{(1-x)}x^2\,{\mathrm ln}x
\eeq
with $x=(1-\beta_\mu)/(1+\beta_\mu)$ and $\beta=(1-4m_\mu^2/s)^{1/2}$ (see 
also remarks concerning the numerical stability of $K(s)$ in 
Ref.~\cite{eidelman}). 
The function $K(s)$ decreases monotonically with increasing $s$. It gives
a strong weight to the low energy part of the integral~(\ref{eq_integral1}).
About 91\pc\ of the total contribution to \amuhad\ is accumulated at c.m. 
energies $\sqrt{s}$ below 2.1~GeV while 72\pc\ of \amuhad\ is covered by 
the two-pion f\/inal state which is dominated by the $\rho(770)$ resonance. 
Recent $\tau$ data published by the ALEPH Collaboration provide a very 
precise spectrum of the two-pion f\/inal state as well as new input for the 
more controversial four-pion f\/inal states. This new information can
signif\/icantly improve the \amuhad\ determination.
%
%
\subsection*{Running of the QED F\/ine Structure Constant}

In the same spirit we evaluate the hadronic contribution \daqed\
to the renormalized vacuum polarization function $\Pi_\gamma^\prime(s)$ 
which governs the running of the electromagnetic f\/ine structure constant 
\aqed. For the spin 1 photon, $\Pi_\gamma^\prime(s)$ is given by the Fourier
transform of the contraction of the electromagnetic currents 
$j_{\mathrm em}^\mu(s)$ in the vacuum
$(q^\mu q^\nu-q^2g^{\mu\nu})\,\Pi_\gamma^\prime(q^2)=i\int d^4x\,e^{iqx}\langle
0|T(j_{\mathrm em}^\mu(x)j_{\mathrm em}^\nu(0))|0\rangle$.  With 
\daqed=$-4\pi\alpha\,{\mathrm Re}
\left[\Pi_\gamma^\prime(s)-\Pi_\gamma^\prime(0)\right]$, one has
\beq
    \alpha(s) \:=\: \frac{\alpha(0)}{1-\Delta\alpha(s)}~,
\eeq
where $4\pi\alpha(0)$ is the square of the electron charge in the 
long-wavelength Thomson limit. The contribution \daqed\ can naturally 
be subdivided in a leptonic and a hadronic part. Furthermore, at 
$s=M_{\mathrm Z}^2$ it is appropriate to separate the leading vacuum 
polarization contribution involving the f\/ive light quarks $u,d,s,c,b$ from 
the top quark contribution since the latter cannot be calculated in the 
light fermion approximation. 

The leading order leptonic contribution is given by
\beq
   \Delta\alpha_{\mathrm lep}(M_{\mathrm Z}^2)=
        \frac{\alpha(0)}{3\pi}\sum_\ell
            \left({\mathrm ln}\frac{s}{m^2_\ell}-\frac{5}{3}\right)
             \:=\: 314.2\times10^{-4}~. 
\eeq
Using analyticity and unitarity, the dispersion integral for the contribution
from the light quark hadronic vacuum polarization \daqedhZ\ reads~\cite{cabbibo}
\beq\label{eq_integral2}
    \Delta\alpha_{\mathrm had}^{(5)}(M_{\mathrm Z}^2) \:=\:
        -\frac{M_{\mathrm Z}^2}{4\pi^2\,\alpha}\,
         {\mathrm Re}\intl_{4m_\pi^2}^{\infty}ds\,
            \frac{\sigma_{\mathrm had}(s)}
                 {s-M_{\mathrm Z}^2-i\epsilon}~,
\eeq
where $\sigma(s)=16\pi^2\alpha^2(s)/s\cdot{\mathrm Im}\Pi_\gamma^\prime(s)$
from the optical theorem, and Im$\Pi_\gamma^\prime$ stands for the 
absorptive part of the hadronic vacuum polarization correlator. In contrast to 
\amuhad, the integration kernel favours cross sections at higher masses. 
Hence, the improvement when including $\tau$ data is expected to be small.

The top quark contribution can be calculated using the next-to-next-to-leading
order $\alpha_s^3$ prediction of the total inclusive cross section ratio $R$, 
def\/ined as
\beq
     R(s) \:=\: \frac{\sigma_{\mathrm tot}(e^+e^-\rightarrow\mathrm hadrons)}
                     {\sigma(e^+e^-\rightarrow\mu^+\mu^-)}
          \:=\: \frac{3s}{4\pi\alpha^2}\sigma_{\mathrm tot}~,
\eeq
from perturbative QCD~\cite{rqcd,eidelman}:
\beq\label{eq_pert}
     R_{\mathrm{pert}}(s) 
        = 3\sum_fQ_f^2 \left(1-\frac{4m_f^2}{s}\right)^{\!\!1/2}\!
                         \left(1+\frac{2m_f^2}{s}\right)
                   \left[
                         1+\frac{\alpha_s}{\pi} + 
                         r_1\left(\frac{\alpha_s}{\pi}\right)^{\!\!2} +
                         r_2\left(\frac{\alpha_s}{\pi}\right)^{\!\!3\,}
                   \right]~,
\eeq
where $r_1=1.9857-0.1153\,n_f$, 
$r_2=-6.6368-1.2001\,n_f-1.2395(\sum_fQ_f)^2/3\sum_fQ_f^2$ and $n_f$ 
is the number of involved quark flavours. The evaluation of the 
integral~(\ref{eq_integral2}) with $m_{\mathrm top}=175$~GeV 
and the running strong coupling constant f\/ixed at 
$\alpha_s(M_{\mathrm Z}^2)=0.121$ yields 
$\Delta\alpha_{\mathrm top}(M_{\mathrm Z}^2)=-0.6\times10^{-4}$.
\vs
Using \daqedhZ$=(280\,\pm\,7)\times10^{-4}$~\cite{eidelman}, one obtains
\beq\label{eq_alpha}
    \alpha^{-1}(M_{\mathrm Z}^2) \:=\: 128.902 \,\pm\, 0.090~.
\eeq
Again, the error is dominated by the hadronic vacuum polarization part 
that is not calculable within perturbative QCD.  

%
%

\section{The Integration Procedure}
\label{sec_integration}

The information used for the evaluation of the integrals~(\ref{eq_integral1}) 
and (\ref{eq_integral2}) comes mainly from direct measurements of the cross
sections in \ee\ annihilation and \via\ CVC from $\tau$ \sfs.
The integrals themselves are evaluated using the trapezoidal rule, {\it i.e.}, 
combining adjacent measurement points by straight lines. Even if this 
method is straightforward and free from theoretical constraints (other 
than CVC in the $\tau$ case), its numerical calculation requires special 
care. The combination of measurements from dif\/ferent experiments taking
into account correlations -- both inside each data set and between
dif\/ferent experiments are subjected to additional studies and discussions. 

%
%
\subsection*{Averaging Data from dif\/ferent Experiments}

In order to exploit the maximum information, we combine weighted 
measurements of dif\/ferent experiments at a given energy instead of 
calculating separately the integrals for every experiment and f\/inally 
averaging them\footnote
{
   One could imagine two experiments $a$ and $b$, each with two independent
   measurements $a_1(E_1)\pm\Delta a_1$, $a_2(E_2)\pm\Delta a_2$ and
   $b_1(E_1)\pm\Delta b_1$, $b_2(E_2)\pm\Delta b_2$ at energies $E_1\neq E_2$.
   Setting $\Delta a_1=\Delta a_2=\Delta b_1=\Delta b_2$ leads to identical
   errors in both integration methods. However, non-symmetric errors
   as, {\it e.g.}, $\Delta a_1/2=2\Delta a_2=2\Delta b_1=
   \Delta b_2/2$ propagate a 53\pc\ larger uncertainty when calculating 
   independently the sum over the points (trivial integration) of the 
   experiments $a$, $b$ and averaging afterwards rather than averaging 
   $\langle a_1,\,b_1\rangle$ and $\langle a_2,\,b_2\rangle$ f\/irst, 
   {\it i.e.}, keeping the energy information of the respective points in 
   the average. \\
}.
Generally, if dif\/ferent measurements at a given c.m. energy show 
inconsistencies, {\it i.e.}, their $\chi^2$ per degree of freedom (dof) 
is larger than one, we rescale the error of their weighted average with 
$\sqrt{\chi^2/\mathrm dof}$.
\vs
The solution of the averaging problem is found using a correlated $\chi^2$ 
minimization. We def\/ine
\beq\label{eq_chi2}
     \chi^2 \:=\: \sum_{n=1}^{N_{\mathrm exp}} 
                      \sum_{i,j=1}^{N_n}
                          (x_i^n-k_i)\,(C_{ij}^n)^{-1}\,(x_j^n-k_j)~,
\eeq
where $x_i^n$ is the $i$th cross section measurement of the $n$th experiment
in a given f\/inal state, $C_{ij}^n$ is the covariance between the $i$th and 
the $j$th measurement and $k_i$ is the unknown distribution. The covariance 
matrix $C^n$ is def\/ined as
\beq\label{eq_cov}
     C_{ij}^n \:=\: \Bigg\{ 
                    \begin{array}{l@{\quad\quad}l}
                      (\Delta_{i,\mathrm stat}^n )^2 +
                        (\Delta_{i,\mathrm sys}^n )^2 ~~~{\mathrm for}~i=j \\
                      \Delta_{i,\mathrm sys}^n \cdot
                        \Delta_{j,\mathrm sys}^n  ~~~{\mathrm for}~i\neq j
                    \end{array} ,~~~i,j=1,\dots,N_n~,
\eeq
where $\Delta_{i,\mathrm stat}^n$ 
($\Delta_{i,\mathrm sys}^n$) denotes the statistical (systematic) error
of $x_i^n$. The systematic errors of the \ee\ annihilation measurements 
are essentially due to luminosity and ef\/f\/iciency uncertainties.
The minimum condition $d\chi^2/dk_i=0,~\forall i$ leads to the linear 
equation problem
\beq\label{eq_dchi2}
   \sum_{n=1}^{N_{\mathrm exp}} 
                      \sum_{j=1}^{N_n}\,
                          (x_j^n-k_j)\,(C_{ij}^n)^{-1} 
                          \:=\: 0~,~~~~~i=1,\dots,N_n~.
\eeq
The inverse covariance $\tilde{C}_{ij}^{-1}$ between the solutions $k_i$, 
$k_j$ is the sum over the inverse covariances of every experiment
\beq\label{eq_ddchi2}
    \tilde{C}_{ij}^{-1} \:=\: \sum_{n=1}^{N_{\mathrm exp}} (C_{ij}^n)^{-1}~.
\eeq
\par
\subsection*{Correlations between Experiments}

Eq.~(\ref{eq_ddchi2}) provides the covariance matrix needed for the 
error propagation when calculating the integrals over the solutions 
$k_i$ from Eq.~(\ref{eq_dchi2}). Up to this point, $\tilde{C}_{ij}$
only contains correlations between the systematic uncertainties within
the same experiment. However, due to commonly used simulation techniques
for acceptance and luminosity determinations as well as state-of-the-art
calculations of radiative corrections, systematic correlations from 
one experiment to another cannot be excluded. It is clearly a 
dif\/ficult task to reasonably estimate the amount of such correlations
as they depend on the capabilities of the contributing experiments and
one's theoretical understanding of the dynamics of the respective f\/inal 
states. In general, one can state that in older
experiments, where only parts of the total solid angle were covered
by the detector acceptance, individual experimental limitations should
dominate the systematic uncertainties. Potentially common systematics, such
as radiative corrections or ef\/ficiency, acceptance and luminosity
calculations based on the Monte Carlo simulation, play only minor roles. 
The correlations between systematic errors below 2~GeV c.m. energy are 
therefore estimated to be between 10\pc\ and 30\pc, with the exception of 
the $\pi^+\pi^-$ f\/inal state, where we impose a 40\pc\ correlation 
due to the easier experimental situation and the better knowledge of 
the dynamics which leads to non-negligible systematic contributions from 
the uncertainties of the radiative corrections. At energies above 2.1~GeV 
the experiments measured the total inclusive cross section ratio $R$. 
Between 2.1 and 3.1~GeV, individual technical problems dominate 
the systematic uncertainties. At higher energies, new experiments provide 
nearly full geometrical acceptance which decreases the uncertainty of 
ef\/ficiency estimations. Radiative corrections as well as theoretical
errors of the luminosity determination give important contributions
to the f\/inal systematic errors quoted by the experiments. We therefore
estimate the correlations between the systematic errors of the experiments 
to be negligible between 2~GeV and 3~GeV, 20\pc\ between 3~GeV and
10~GeV, and 30\pc\ above 10~GeV. 

These correlation coef\/f\/icients are added to all those entries of 
$\tilde{C}_{ij}$ from Eq.~(\ref{eq_ddchi2}) which involve two dif\/ferent 
experiments. 

%
%
\subsection*{Inclusion of \boldmath$\tau$ Data}

In this analysis we include additional data from $\tau$ decays into two-
and four-pion f\/inal states\footnote
{
  Throughout this paper, charge conjugate states are implied.
}: 
\tauto\pipiz\nut, \tauto\pitpiz\nut\ and \tauto\tpipiz\nut, 
recently published by the ALEPH Collaboration~\cite{Aleph}.
Assuming isospin invariance to hold, the corresponding \ee\ isovector
cross sections are calculated \via\ the CVC relations
\begin{eqnarray}
\label{eq_cvc_2pi}
 \sigma_{e^+e^-\rightarrow\,\pi^+\pi^-}^{I=1}
        & \:=\: &
 \frac{4\pi\alpha^2}{s}\,v_{1,\,\pi^-\pi^0\,\nu_\tau}~, \\[0.3cm]
\label{eq_cvc_4pi}
 \sigma_{e^+e^-\rightarrow\,\pi^+\pi^-\pi^+\pi^-}^{I=1} 
        & \:=\: &
             2\cdot\frac{4\pi\alpha^2}{s}\,
             v_{1,\,\pi^-\,3\pi^0\,\nu_\tau}~, \\[0.3cm]
\label{eq_cvc_2pi2pi0}
 \sigma_{e^+e^-\rightarrow\,\pi^+\pi^-\pi^0\pi^0}^{I=1} 
        & \:=\: &
             \frac{4\pi\alpha^2}{s}\,
             \left[v_{1,\,2\pi^-\pi^+\pi^0\,\nu_\tau} 
                  \:-\:
                     v_{1,\,\pi^-\,3\pi^0\,\nu_\tau}
             \right]~.
\end{eqnarray}
The $\tau$ \sfs\ $v_1$ are given as binned continous distributions of the 
mass-squared $s$. In each bin $i$, the \sf\ $v_1(s_i)$ contains the
(normalized) invariant mass spectrum $\Delta N_i/N$, integrated over
the bin width $ds_i$. On the contrary \ee\ cross sections are 
measured at discrete energy settings. To each \ee\ measurement
is associated a $\tau$ cross section value obtained by interpolating
between adjacent bins. This interpolation takes into account 
the correlations between the bins and is achieved imposing
a functional form obtained from a f\/it
to Breit-Wigner resonances~\cite{Aleph} using the Gounaris-Sakurai
parametrization~\cite{gounarissak}. However, the f\/it function is 
renormalized in each bin so that its integral over the width of each bin
corresponds to the measured CVC cross section for that bin.
All the data points -- $\tau$, \ee\ and interpolated $\tau$ values --
are injected with their correlations into Eqs.~(\ref{eq_dchi2}) and 
(\ref{eq_ddchi2}).
\vs
Due to the high bin-to-bin correlations of the
$\tau$ data and the signif\/icant normalization uncertainty coming from 
the $\tau$ hadronic branching ratios, biases of the least-square 
minimization~\cite{Agostini} cannot be excluded. We therefore calculate 
the average solution twice, {\it i.e.}, with and without correlations,
use the mean value of both integrations as the result and add half of the 
total dif\/ference as additional systematic error. This is done in all 
cases where $\tau$ data are involved. The ef\/fect amounts to about 
10\pc\ of the total error.

%
%
\subsection*{Evaluation of the Integral}

The procedure described above provides the weighted average and the covariance 
of the cross sections from dif\/ferent experiments contributing to a 
certain f\/inal state in a given range of c.m. energies. We now apply the 
trapezoidal rule. In order to perform the integrations~(\ref{eq_integral1}) 
and (\ref{eq_integral2}), we subdivide the integration range in 
suf\/f\/iciently small energy steps and calculate for each of these steps 
the corresponding covariance (where additional correlations induced by 
the trapezoidal rule has to be taken into account). This procedure yields
error envelopes between adjacent measurements as depicted by the shaded
bands in Figs.~\ref{fig_2pi} and \ref{fig_sf_vgl_rest}.

%
%
\section{Radiative Corrections}

Higher order radiative corrections bias the measurements of both cross 
sections in \ee\ annihilation and invariant mass spectra from $\tau$ 
hadronic decays. The \ee\ experiments generally correct the measured cross 
section for QED ef\/fects, including bremsstrahlung, vacuum polarization and 
higher order self-energy graphs (see references in~\cite{eidelman}). 
Following the prescription of Ref.~\cite{eidelman}, we have multiplied all 
inclusive cross section measurements $R$ at masses below the $J/\psi$ 
resonance by the (small) correction factor 
$(1+\Delta\alpha_{\mathrm{lep}}(s))(\alpha/\alpha(s))^2$ 
in order to account for the missing correction for hadronic vacuum 
polarization.
\vs
In $\tau$ decays, 
f\/inal state radiation from the $\tau$ itself or from its decay 
products can inf\/luence the invariant mass measurement. Due to the high 
mass of the $\tau$ lepton, the bremsstrahlung graph is largely suppressed. 
Both types of radiation are included in the Monte Carlo simulation program 
KORALZ~\cite{koralz}, used by ALEPH to unfold the measured 
distributions from detector resolution and physical (higher order) ef\/fects. 
Even if the actual frequency with which f\/inal state radiation occurs or if
its energy was not well simulated in the Monte Carlo, reconstructed photons
found to originate from radiation of the $\tau$ decay~\cite{aleph_hbr} are 
included in the invariant mass determination, and thus do not bias the 
measurement.
\vs
Electroweak radiative corrections are applied through the CVC correction
factor $S_{\mathrm EW}$ def\/ined in Eq.~(\ref{eq_sew}).

%
%
\section{The Origin of the Data}

The exclusive low energy \ee\ cross sections have mainly been measured by 
experiments working at \ee\ colliders in Novosibirsk and Orsay. Due to the 
high hadron multiplicity at energies above 2.5$--$3.1~GeV, the exclusive 
measurement of the respective hadronic f\/inal states is not practicable.
Consequently, the experiments at the high energy colliders DORIS and PETRA  
(DESY) and PEP (SLAC) have measured the total inclusive cross section ratio 
$R$.
\vs
We give in the following a compilation of the data used in this analysis:
\begin{itemize}
\item[--] The \ee$\rightarrow\pi^+\pi^-$ measurements are taken from 
  OLYA~\cite{barkov,E_54}, TOF~\cite{E_55}, NA7~\cite{E_56}, CMD~\cite{barkov}, 
  DM1~\cite{E_58} and DM2~\cite{E_59}. In addition, we use $\tau$ data 
  from ALEPH~\cite{Aleph} normalized to the world average branching ratio 
  B(\tauto\pipiz\nut)\,=\,(25.24\pms0.16)\pc~\cite{PDG}. According to
  Eq.~(\ref{eq_cvc_2pi}), $\tau$ data provide only the dominant isovector 
  part of the total two-pion cross section. A correction due to the small 
  isospin-violating isoscalar $\omega\rightarrow\pi^+\pi^-$ f\/inal state,
  which interferes with the isovector amplitude, is applied. A small 
  correction for the missing, {\it i.e.}, unmeasured decay modes 
  $\rho\rightarrow\pi^0\gamma$ (only for \ee\ data) and 
  $\rho\rightarrow\;\eta\gamma$, is added.
\item[--] The reaction \ee$\rightarrow\pi^+\pi^-\pi^0$ is dominated by $\omega$ 
  and $\phi$ intermediate resonances. In the peak region of these resonances
  we use analytic parametrizations of the cross sections. The non-resonant
  data are taken from ND~\cite{E_64}, M2N~\cite{E_65}, M3N~\cite{E_66},
  DM1~\cite{E_68} and DM2~\cite{E_69}. Corrections for the missing
  $\omega$ and $\phi$ decay modes are applied.
\item[--] The \ee$\rightarrow\pi^+\pi^-\pi^0\pi^0$ data are available from 
  OLYA~\cite{E_70}, ND~\cite{E_64}, M2N~\cite{E_65}, 
  DM2~\cite{E_74,schioppa,E_74p,E_75} and M3N~\cite{E_66}. According to 
  Eq.~(\ref{eq_cvc_2pi2pi0}), a linear combination of both four-pion 
  $\tau$ decay channels measured by ALEPH~\cite{Aleph} connects the 
  corresponding \sfs\ with the above \ee\ f\/inal state. We use the two
  branching ratios B(\tauto\tpipiz\nut)\,=\,(4.25\pms0.09)\pc\ and 
  B(\tauto\pitpiz\nut)\,=\,(1.14\pms0.14)\pc~\cite{PDG}, as an
  appropriate normalization.
\item[--] The reaction \ee$\rightarrow\omega\pi^0$ is mainly reconstructed in 
  the $\pi^+\pi^-\pi^0\pi^0$ f\/inal state. It was studied by the 
  collaborations ND~\cite{E_64} and DM2~\cite{E_74p}. Corrections for the 
  missing $\omega$ decay modes are applied.
\item[--] The \ee$\rightarrow\pi^+\pi^-\pi^+\pi^-$ f\/inal state was studied 
  by the experiments OLYA~\cite{E_70}, ND~\cite{E_64}, MEA~\cite{E_67}, 
  CMD~\cite{E_72}, DM1~\cite{E_73,PhysLett81B}, DM2~\cite{E_74,E_74p,E_75} 
  and M3N~\cite{E_66}. The corresponding \sf\ from \tauto\pitpiz\nut\ 
  (according to Eq.~(\ref{eq_cvc_4pi})) is provided by ALEPH~\cite{Aleph}.
\item[--] The \ee$\rightarrow\pi^+\pi^-\pi^+\pi^-\pi^0$ f\/inal state is taken 
  from M3N~\cite{E_66} and CMD~\cite{E_72}. The other f\/ive-pion mode
  \ee$\rightarrow\pi^+\pi^-3\pi^0$ can be accounted for using the rigourous 
  isospin relation
  $\sigma_{\pi^+\pi^-3\pi^0}=0.5\times\sigma_{\pi^+\pi^-\pi^+\pi^-\pi^0}$.
\item[--] For the reaction \ee$\rightarrow\omega\pi^+\pi^-$, measured by the 
  groups DM1~\cite{E_73} and DM2~\cite{E_72}, a correction for $\omega$ decays
  other than into three pions which appear in the f\/ive-pion f\/inal state 
  is applied.
\item[--] The \ee$\rightarrow\pi^+\pi^-\eta$ data were studied by 
  ND~\cite{E_64} and DM2~\cite{E_80}. We subtract from the cross section the 
  contributions which are already counted in the $\pi^+\pi^-\pi^+\pi^-\pi^0$ 
  and $\pi^+\pi^-3\pi^0$ f\/inale states.
\item[--] The cross sections of the six-pion f\/inal states $3\pi^+3\pi^-$ 
  and $2\pi^+2\pi^-2\pi^0$ were measured by DM1~\cite{E_79}, M3N~\cite{E_66}
  CMD~\cite{E_72} and DM2~\cite{schioppa}. In Ref.~\cite{Aleph} an upper 
  limit for the unknown $\pi^-\pi^+\,4\pi^0$ cross section of 
  $\sigma_{\pi^+\pi-4\pi^0}\le(3/2)\times\sigma_{2\pi^-2\pi^+\,2\pi^0}
  -(9/24)\times\sigma_{3\pi^-3\pi^+}$ was derived using isospin constraints. 
  We take half of this upper limit as the estimated contribution, with an 
  error of 100\pc.
\item[--] The \ee$\rightarrow$\,K$^+$K$^-$ and \ee$\rightarrow$\,\Ks\Kl\ 
  cross sections are taken from OLYA~\cite{E_82}, DM1~\cite{LAL80_xx} and 
  DM2~\cite{E_87}.
\item[--] The reactions \ee$\rightarrow$\,\Ks\,K$^+\pi^-$ and  
  \ee$\rightarrow$\,K$^+$K$^-\pi^0$ were studied by DM1~\cite{E_88a,E_88b} 
  and DM2~\cite{E_74}. Using isospin symmetry the cross 
  section of the f\/inal state \Ks\Kl\piz\ is obtained from the relation 
  $\sigma_{{\mathrm K}_{\mathrm S}^0{\mathrm K}_{\mathrm L}^0\pi^0}
  =\sigma_{{\mathrm K}^+{\mathrm K}^-\pi^0}$.
\item[--] The inclusive reaction \ee$\rightarrow$\,\Ks+$X$ was analyzed by 
  DM1~\cite{DM1thesis}. After subtracting from its cross section the separately
  measured contributions of the f\/inal states \Ks\Kl, \Ks\,K$^+\pi^-$ and 
  \Ks\Kl\piz, it still includes the modes \Ks\Ks$\pi^+\pi^-$, 
  \Ks\Kl$\pi^+\pi^-$, \Ks${\mathrm K}^+\pi^-\pi^0$ and 
  \Ks${\mathrm K}^-\pi^+\pi^0$.
  With the assumption that the cross sections for the processes 
  \ee$\,\rightarrow$K$^0\bar{\mathrm K}^0(\pi\pi)^0$ and
  \ee$\,\rightarrow$K$^+{\mathrm K}^-(\pi\pi)^0$ are equal, one can
  summarize the total K${\mathrm\bar{K}}\pi\pi$ contribution as twice 
  the above corrected \Ks+$X$ cross section. A reasonable
  estimate of the systematic uncertainty, implied by the assumption 
  made, is obtained by taking the cross section for the channel 
  K$^+$K$^-\pi^+\pi^-$ measured by DM1~\cite{E_88a} 
  and DM2~\cite{E_74}.
\item[--] At higher energy the total cross section ratio $R$ is measured 
  inclusively. We use the data provided by the experiments 
  $\gamma\gamma2$~\cite{E_78}, MARK~I~\cite{E_96}, DELCO~\cite{delco}, 
  DASP~\cite{dasp}, PLUTO~\cite{pluto}, LENA~\cite{lena}, Crystal Ball~\cite{CB},
  MD-1~\cite{MD1}, CELLO~\cite{cello}, JADE~\cite{jade}, MARK-J~\cite{markj},
  TASSO~\cite{tasso}, CLEO~\cite{cleo}, CUSB~\cite{cusb} and MAC~\cite{mac}. 
  Above 3.5~GeV the measurements of the MARK~I Collaboration are signif\/icantly 
  higher than those from LENA, PLUTO and Crystal Ball.
  In addition, the QCD prediction of $R$, which should be reliable in this 
  energy regime, favours lower values. In agreement with Ref.~\cite{eidelman},
  we neglect MARK~I data above this energy threshold. \par
  \hspace{0.5cm}Although small, the enhancement of the cross section due to 
  $\gamma$--Z interference is corrected for c.m. energies above the $J/\psi$ 
  mass. We use a factorial ansatz according to Ref.~\cite{burkjeger,eidelman},
  yielding a negligible contribution to \amuhad\ and a $-0.30\times10^{-4}$ 
  shift of \daqedhZ.
\end{itemize}

%
%
\section{Analytical Contributions}

In some energy regions where data information is scarce and/or reliable 
theoretical predictions are available, we use analytical contributions
to extend the experimental integral.

%
%
\subsection*{The \boldmath$\pi^+\pi^-$ Threshold Region}

To overcome the lack of data at threshold energies, a second order 
expansion obtained from {\it Chiral Perturbation Theory}~\cite{chiral_exp} 
is used as a description of the pion form factor $F_\pi$ (which is 
connected with the two-pion cross section \via\ the expression 
$|F_\pi|^2=3s\sigma_{\pi\pi}/(\pi\alpha^2(1-4m_\pi^2/s)^{3/2})$:
\beq\label{eq_chir_exp}
F_{\pi}^{\mathrm ChPT} \;\simeq\; 
      1 + \frac{1}{6}\langle r^2 \rangle_\pi\,s + c_\pi\,s^2 +
      O(s^3)~.
\eeq
Exploiting precise results from space-like data~\cite{space_like}, the pion 
charge radius-squared $\langle r^2 \rangle_\pi=(0.431\,\pm\,0.026)$~fm$^2$ 
and the coef\/f\/icient $c_\pi=(3.2\,\pm\, 1.0)$~GeV$^{-4}$ from 
Eq.~(\ref{eq_chir_exp}) have recently been determined~\cite{finkemeier} 
by means of a simultaneous f\/it.

%
%
\subsection*{Narrow Resonances}

The \ee\ annihilation cross section involves narrow resonances 
such as the $\omega(782)$ and $\phi(1020)$ at low energies,
the $J/\psi$ and $\Upsilon$ resonances at the c$\mathrm\bar{c}$
and b$\mathrm\bar{b}$ quark thresholds, respectively, as well as their 
excited spectroscopic states. It is safe to parametrize these states using
relativistic Breit-Wigner resonance shapes with an $s$-dependent width. We 
use the formulae given in Ref.~\cite{eidelman}. The physical input values of
the parametrizations and their errors are taken from Ref.~\cite{PDG}. The
total parametrization errors are then calculated by gaussian error 
propagation.
%
%
\subsection*{High Energy Tail}

At energies suf\/f\/iciently above the $\Upsilon$ resonance family, the 
perturbative QCD prediction of $R$ with five active quarks is 
supposed to be reliable. In agreement with Ref.~\cite{eidelman} we use 
$R_{\mathrm{pert}}(s)$ from Eq.~(\ref{eq_pert}) for $\sqrt{s}\ge40$~GeV.

%
%
\section{Results}
\label{sec_results}

We evaluate the integrals~(\ref{eq_integral1}) and (\ref{eq_integral2})
exclusively, {\it i.e.}, for every contributing f\/inal state, up to the
c.m. energy of 2.125~GeV. Even if some particular modes have been measured 
up to somewhat higher energies,
we have to worry about unmeasured exclusive modes and therefore use the
total $R$ measurement above this threshold. Both energy regions are assumed
to be uncorrelated. Because the contributions of the exclusive channels 
at low energy are simply summed up, we have to estimate their respective
covariances when propagating the error: in general, unmeasured f\/inal states 
whose contributions are deduced from measured ones \via\ isospin are set 
to be 100\pc\ correlated with these. Also
dif\/ferent detectors performing the same measurement are correlated 
through the sharing of commonly used simulation techniques to calculate 
acceptance and selection ef\/ficiency which depend on the assumed underlying 
physical dynamics. Contributions from resonances that are analytic are 
globally assumed to have 20\pc\ correlations due to modelling uncertainties.
Between purely measured f\/inal states we have estimated the correlations
depending on the number of common experiments that contribute to their
measurements and on the common energy region, as well as according to the
relative importance of their statistical and systematic errors. In general 
our estimation yields a correlation between 10\pc\ and 20\pc. 
This treatment is dif\/ferent from that of Ref.~\cite{worstell} where
a 100\pc\ correlation was assumed. 
\vs
As described in Section~\ref{sec_cvc}, corrections to the charged
$\rho^\pm$ width have to be applied to account for small CVC-violating 
ef\/fects. The magnitude of the width dif\/ference Eq.~(\ref{eq_gamdif})
translated into $a_\mu^{\rm had}$ and \daqedhZ\ is evaluated using a 
parametrization of the $\rho$ line shape based on vector 
resonances~\cite{Aleph}. We obtain the additive corrections
\beqn
\label{eq_adif}
   \delta a_\mu^{\rm had} 
       &=& 
         -(1.3 \pm 2.0)\times10^{-10} \nonumber\\
   \delta \Delta\alpha_{\rm had}^{(5)}(M_{\rm Z}^2) 
       &=&
         -(0.09 \pm 0.12)\times10^{-4}
\eeqn
for the \tauto\pipiz\nut\ \sf\ which is applied in the present analysis. 
Corrections from the higher mass resonances $\rho(1450), \rho(1700)$ 
are expected to be negligible. 
\vs
\begin{figure}[p]
\begin{center}
\mbox{
\epsfig{file=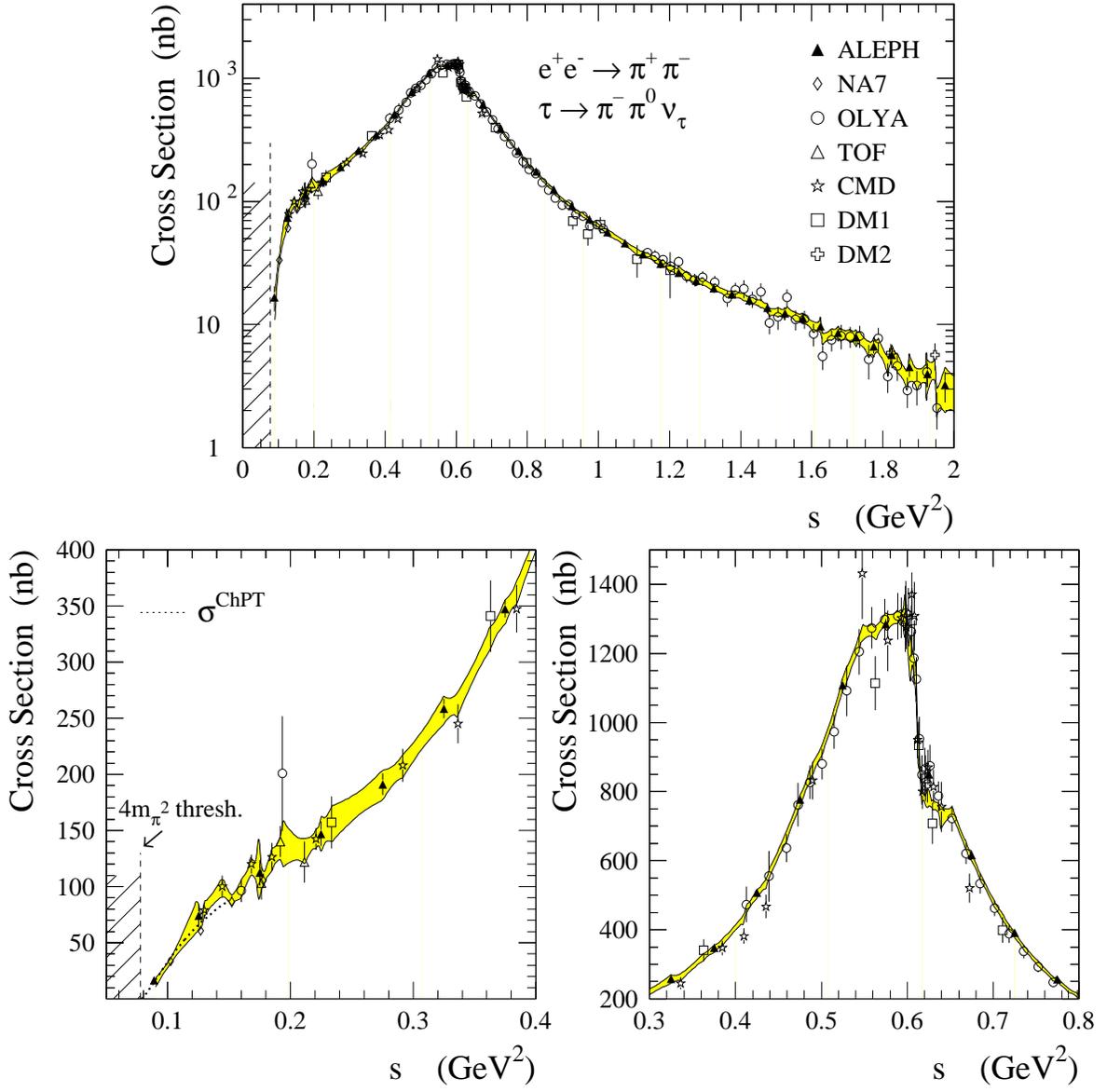,height=15.4cm,width=15.4cm,bbllx=20pt,bblly=20pt,
bburx=560pt,bbury=560pt}}
\end{center}
\caption[.]{\it Two-pion cross section as a function of the c.m.
             energy-squared. The band represents the result of the averaging 
             procedure described in the text within its diagonal errors. 
             The lower left hand plot shows
             the chiral expansion of the two-pion cross section obtained 
	     from expression~(\ref{eq_chir_exp}).}
\label{fig_2pi}
\end{figure}
\begin{figure}[thb]
\begin{center}
\mbox{
\epsfig{file=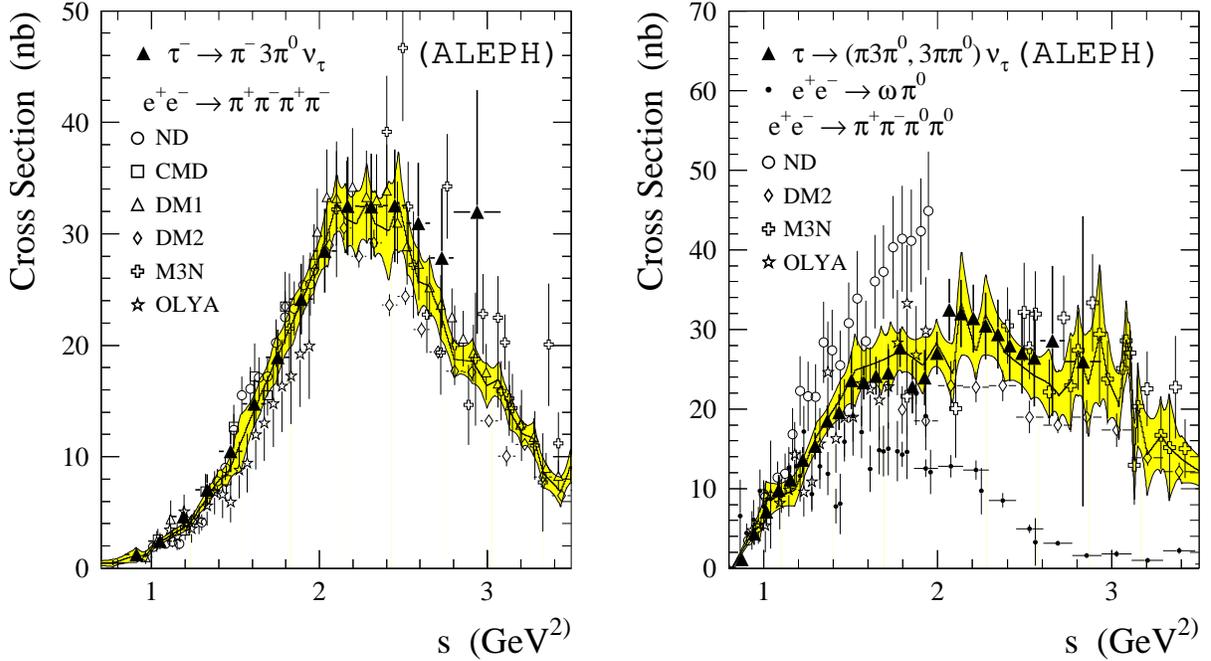,height=8.5cm,bbllx=20pt,bblly=20pt,
bburx=570pt,bbury=320pt}}
\end{center}
\caption{\it Four-pion cross section as a function of the c.m. 
             energy-squared. The band represents the result of the averaging 
             procedure described in the text within its diagonal errors.
             The right-hand plot shows additionally the 
             \ee$\rightarrow\omega\pi^0$ amplitude (small points).}
\label{fig_sf_vgl_rest}
\end{figure}
\begin{table}[p]
\begin{center}
{\small
\begin{tabular}{ |c|c|c|c| } \hline 
F\/inal states      & \amuhad\ ($\times10^{10}$)  & \daqedhZ\ ($\times10^{4}$) 
                    & Energy (GeV) \\ \hline\hline
 $\pi^+\pi^-$ threshold  
                    &   2.30 $\pm$  0.05 &  0.04 $\pm$ 0.00 & $4m_\pi^2$\,--\,0.320
\\ \hline
$\pi^+\pi^-$        
                    & 495.86 $\pm$ 12.46 & 34.01 $\pm$ 0.87 & 0.320\,--\,2.125 
\\
$\pi^+\pi^-$ (incl. $\tau$ data)       
                    & 500.81 $\pm$  6.03 & 34.31 $\pm$ 0.38 & 0.320\,--\,2.125 
\\
$\rho(\pi^0\gamma+\eta\,\gamma)$\,$^(\footnotemark[1]{^)}$
                    &   0.30 $\pm$  0.05 &  0.02 $\pm$ 0.01 & 0.298\,--\,2.125 
\\ \hline
$\omega$
                    &  37.09 $\pm$  1.07 &  2.97 $\pm$ 0.09 & 0.420\,--\,0.810
\\
$\omega\rightarrow\pi\gamma$, neutrals\,$^(\footnotemark[1]{^)}$
                    &   0.03 $\pm$  0.01 &   $<0.01$        & 0.810\,--\,2.125  
\\
$\Phi$            
                    &  39.23 $\pm$  0.94 &  5.18 $\pm$ 0.12 & 1.000\,--\,1.055  
\\
$\Phi\rightarrow\eta\,\gamma$, $\pi^0\gamma$\,$^(\footnotemark[1]{^)}$
                    &   0.09 $\pm$  0.01 &  0.01 $\pm$ 0.00 & 1.055\,--\,2.125  
\\ \hline
$\pi^+\pi^-\pi^0$ (below $\Phi$)
                    &   4.12 $\pm$  0.41 &  0.42 $\pm$ 0.04 & 0.810\,--\,1.000 
\\ 
$\pi^+\pi^-\pi^0$ (above $\Phi$)
                    &   1.90 $\pm$  0.72 &  0.46 $\pm$ 0.26 & 1.055\,--\,2.125 
\\ \hline
$\pi^+\pi^-2\pi^0$
                    &  21.41 $\pm$  2.36 &  5.82 $\pm$ 0.63 & 0.910\,--\,2.125 
\\
$\pi^+\pi^-2\pi^0$ (incl. $\tau$ data)
                    &  22.26 $\pm$  1.53 &  6.16 $\pm$ 0.49 & 0.897\,--\,2.125 
\\ \hline
$\omega\,\pi^0(\omega\rightarrow\pi\gamma$, neutr.)\,$^(\footnotemark[1]{^)}$
                    &   0.88 $\pm$  0.11 &  0.18 $\pm$ 0.02 & 0.930\,--\,2.125 
\\ \hline
$\pi^+\pi^-\pi^+\pi^-$
                    &  15.90 $\pm$  1.34 &  4.61 $\pm$ 0.39 & 0.983\,--\,2.125 
\\
$\pi^+\pi^-\pi^+\pi^-$ (incl. $\tau$ data)
                    &  16.50 $\pm$  0.98 &  4.76 $\pm$ 0.31 & 0.794\,--\,2.125 
\\ \hline
$\pi^+\pi^-\pi^+\pi^-\pi^0$
                    &   4.02 $\pm$  0.51 &  1.51 $\pm$ 0.20 & 1.019\,--\,2.125
\\
$\pi^+\pi^-3\pi^0$\,$^(\footnotemark[2]{^)}$
                    &   2.01 $\pm$  0.26 &  0.75 $\pm$ 0.10 & 1.019\,--\,2.125 
\\ \hline
$\omega\,\pi^+\pi^-(\omega\rightarrow\pi\gamma$, neutr.)\,$^(\footnotemark[1]{^)}$
                    &   0.07 $\pm$  0.02 &  0.03 $\pm$ 0.01 & 1.340\,--\,2.125 
\\ \hline
$\pi^+\pi^-\pi^+\pi^-\pi^+\pi^-$
                    &   0.47 $\pm$  0.14 &  0.19 $\pm$ 0.04 & 1.350\,--\,2.125 
\\ 
$\pi^+\pi^-\pi^+\pi^-2\pi^0$
                    &   3.32 $\pm$  0.36 &  1.35 $\pm$ 0.14 & 1.350\,--\,2.125 
\\
$\pi^+\pi^-4\pi^0$\,$^(\footnotemark[2]{^)}$
                    &   2.40 $\pm$  2.40 &  0.98 $\pm$ 0.98 & 1.350\,--\,2.125 
\\ \hline
$\eta\,\pi^+\pi^-$\,$^(\footnotemark[3]{^)}$
                    &   0.51 $\pm$  0.14 &  0.16 $\pm$ 0.05 & 1.075\,--\,2.125 
\\ \hline
K$^+$K$^-$ 
                    &   4.30 $\pm$  0.58 &  0.85 $\pm$ 0.10 & 1.055\,--\,2.055 
\\
\Ks\Kl 
                    &   1.20 $\pm$  0.42 &  0.23 $\pm$ 0.08 & 1.090\,--\,2.125 
\\ \hline
\Ks K$^+\pi^-$ (+ \Kl K$^-\pi^+$\,$^(\footnotemark[2]{^)}$)
                    &   2.04 $\pm$  0.36 &  0.70 $\pm$ 0.12 & 1.340\,--\,2.125 
\\
K$^+$K$^-\pi^0$
                    &   0.42 $\pm$  0.29 &  0.15 $\pm$ 0.10 & 1.440\,--\,2.125 
\\
\Ks\Kl$\pi^0$\,$^(\footnotemark[2]{^)}$
                    &   0.42 $\pm$  0.29 &  0.15 $\pm$ 0.10 & 1.440\,--\,2.125 
\\ \hline
K${\mathrm \bar{K}}\pi\pi$ (all modes)
                    &   4.52 $\pm$  1.65 &  1.82 $\pm$ 0.66 & 1.441\,--\,2.125 
\\ \hline
$J/\psi$(1S,2S,3770)
                    &   8.04 $\pm$  0.52 &  9.97 $\pm$ 0.68 & 3.096\,--\,3.800
\\
$\Upsilon$(1S,2S,3S,4S,10860,11020)
                    &   0.10 $\pm$  0.01 &  1.18 $\pm$ 0.08 & 9.460\,--\,11.20
\\ \hline
$R$             
                    &  41.64 $\pm$  3.61 &164.31 $\pm$ 5.59 &  2.125\,--\,40.0
\\ \hline
$R$ (perturbative)\,$^(\footnotemark[4]{^)}$ 
                    &   0.16 $\pm$  0.00 & 42.82 $\pm$ 0.10 & 40.0\,--\,$\infty$ 
\\ \hline\hline
$\sum\;(e^+e^-\rightarrow\,$hadrons)
                    & 695.0  $\pm$ 15.0  &280.9  $\pm$ 6.3  & $4m_\pi^2\,$--$\,\infty$ 
\\
$\sum\;(e^+e^-\rightarrow\,$hadrons) (incl. $\tau$ data)
                    & 701.1  $\pm$  9.4  &281.7  $\pm$ 6.2  & $4m_\pi^2\,$--$\,\infty$ 
\\ \hline
\end{tabular}
}
\end{center}
{\footnotesize 
\begin{quote}
$^{1}\,$Correction for missing modes (see text). \\ \noindent
$^{2}\,$Deduced from isospin relations (see text). \\ \noindent
$^{3}\,$Without contribution from $\eta\rightarrow\pi^+\pi^-\pi^0$ and
        $\eta\rightarrow3\pi^0$. \\ \noindent
$^{4}\,$Values are taken from~\cite{eidelman}.
\end{quote}
} 
\vspace{-0.3cm}
\label{tab_results}
\caption{\it Summary of the \amuhad\ and \daqedhZ\ contributions from \ee\ 
         annihilation and $\tau$ decays. The line ``$\pi^+\pi^-$ threshold''
         contains the results from the integral over 
         expression~(\ref{eq_chir_exp}) at threshold energies.}
\end{table} 
The two- and four-pion cross sections (incl. the $\tau$ contribution)
in dif\/ferent energy regions are depicted in F\/igs.~\ref{fig_2pi}
and \ref{fig_sf_vgl_rest}. 
The bands are the results within (diagonal) error-envelopes of the 
averaging procedure and the application of the trapezoidal rule 
described in Section~\ref{sec_integration}.

%
%
\subsection*{Lowest Order Hadronic Contributions}

The results of the exclusive contributions to \amuhad\ and \daqedhZ\ are 
presented in Table~\ref{tab_results}. After the inclusion of the $\tau$ 
data, the error of \amuhad\ is dominated much less by the uncertainty
of the two-pion contribution. 

Other important sources are the contradictory $\pi^+\pi^-2\pi^0$ data, as 
well as the unmeasured $\pi^+\pi^-4\pi^0$ f\/inal state.
\begin{figure}[th]
\epsfxsize14cm
\centerline{\epsffile{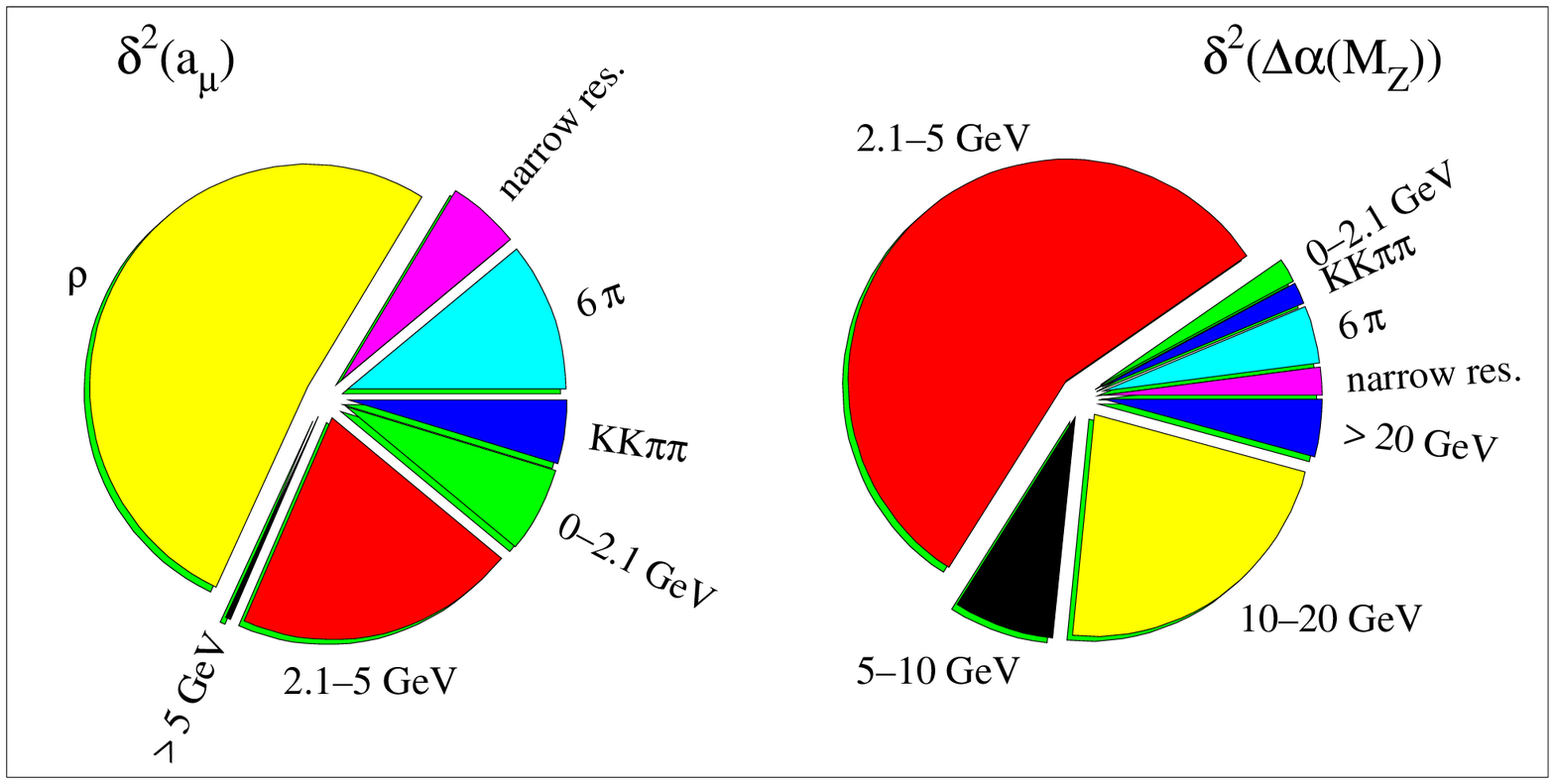}}
\caption[.]{\it Quadratic contribution of the various error sources to
             \amuhad\ (left hand plot) and \daqedhZ\ (right hand plot) after
             the inclusion of $\tau$ data. In the energy region 0--2.1~GeV
             we include all exclusive contributions that are not given
             separately.}
\label{fig_pie}
\end{figure}
In the latter case, limits can be set only by using very conservative
isospin arguments. As shown in Ref.~\cite{Aleph} by means of a decomposition 
in orthogonal isospin classes (Pais~\cite{pais}) the large upper limit comes 
from the assumption of a dominant $\sigma_{411}$ class accompanied by a 
vanishing $\sigma_{321}$ contribution. Both classes occur in 
$\pi^+\pi^-\pi^+\pi^-2\pi^0$, while none contributes to 
$\pi^+\pi^-\pi^+\pi^-\pi^+\pi^-$ and only $\sigma_{411}$ is part of 
$\pi^+\pi^-4\pi^0$. The measured cross section of 
$\pi^+\pi^-\pi^+\pi^-2\pi^0$ is clearly higher 
than the corresponding $\pi^+\pi^-\pi^+\pi^-\pi^+\pi^-$ f\/inal state
hence guaranteeing a leading contribution from one of the classes mentioned
if isospin invariance holds. Since those classes correspond to eigenstates,
a resonance analysis of the measured six-pion data would reveal
important properties of the class structure of the respective modes
which thus could give more constraining isospin bounds. 

Another large uncertainty comes from the K${\mathrm \bar{K}}\pi\pi$ 
f\/inal states. The measurement of the K$^+$K$^-\pi^+\pi^-$ mode alone 
does not allow one to calculate isospin bounds for all possible contributions. 
Fortunately, it is possible to extract the complete K${\mathrm \bar{K}}\pi\pi$
contribution on the basis of a DM1 measurement of the inclusive channel 
\Ks$+X$~\cite{DM1thesis}. Nevertheless, large experimental uncertainties prevent
a precise determination of the corresponding integrals. 

The last important error source, especially for \daqedhZ, comes from the 
integral over the measured high energy inclusive cross section ratio $R$. 
The reliability of the QCD perturbative expansion for 
energies suf\/f\/iciently above the still unpredictable resonance phenomena 
has been proven in many cases (see, {\it e.g.}, $\alpha_s$ measurements from 
dif\/ferent energy scales at LEP, HERA and from $\tau$ decays). Thus, 
theoretical input at energies lower than 40~GeV should be reliable and 
could signif\/icantly help to reduce the integration 
uncertainties~\cite{martin,adel}. However, this has not been used in the 
present analysis which relies on experimental data as far as possible.

\begin{figure}[th]
\epsfxsize17cm
\centerline{\epsffile{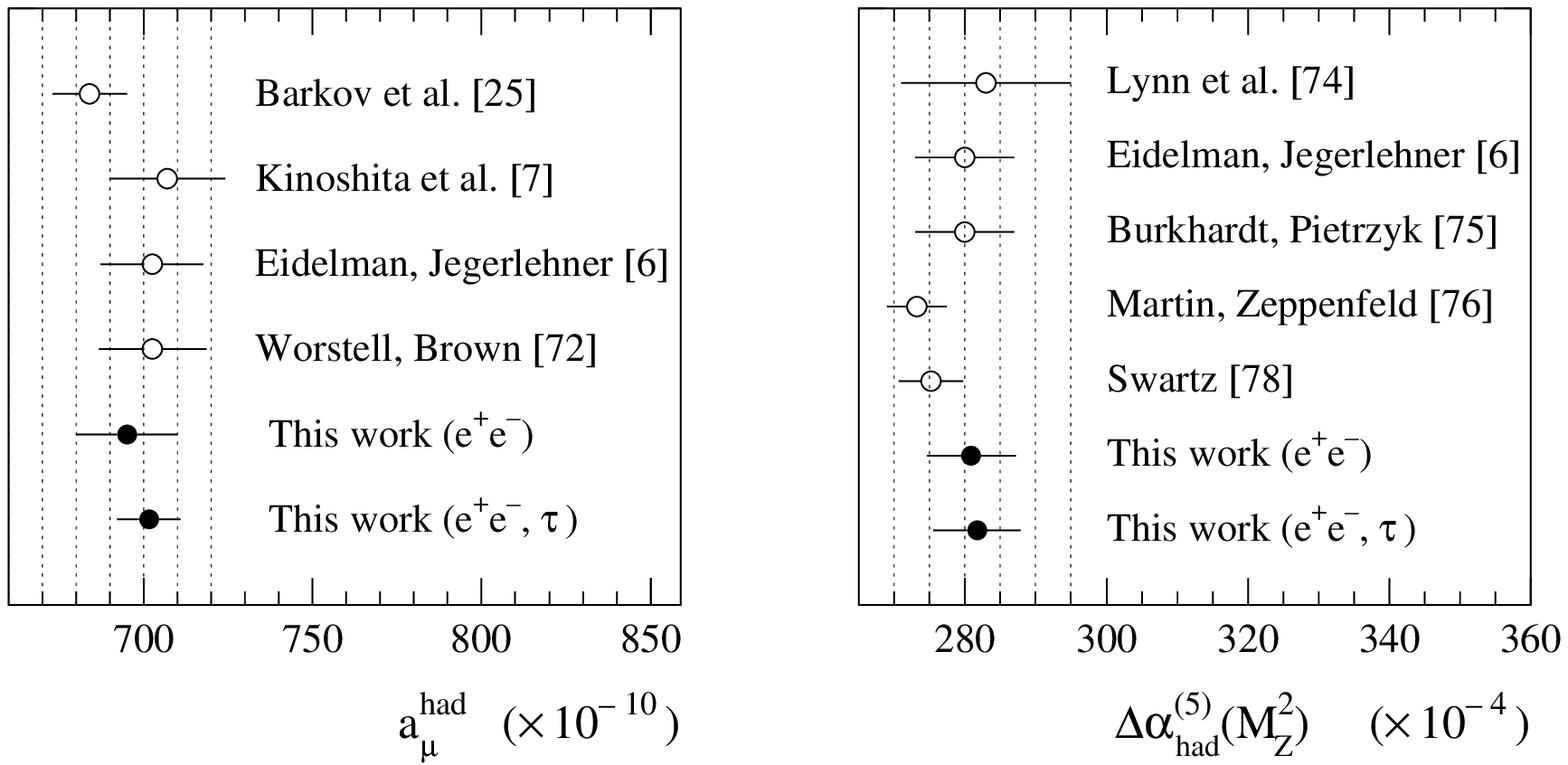}}
\caption[.]{\it Comparison of estimates of \amuhad\ (lowest) and 
            \daqedhZ.}
\label{fig_results}
\end{figure}
The squared contributions of the various f\/inal states and energy regimes
to the errors of \amuhad\ and \daqedZ\ are depicted in Fig.~\ref{fig_pie}. 
Only the results after the inclusion of $\tau$ data are shown. 
\vs
We obtain for the lowest order hadronic vacuum polarization diagram
of the muonic anomalous magnetic moment the contributions
\beqn
     a_\mu^{\mathrm had} & = & (695.0\,\pm\,15.0)\times 10^{-10}
        ~~~~(e^+e^-~\mathrm{data~only}) \nonumber \\
     a_\mu^{\mathrm had} & = & (701.1\,\pm\, 9.4)\times 10^{-10}
        ~~~~({\mathrm{combined}}~e^+e^-~\mathrm{and}~\tau~\mathrm{data})\nonumber
\eeqn
and for the running of $\alpha$ at $M_{\mathrm Z}^2$
\beqn
     \Delta\alpha_{\mathrm had}^{(5)}(M_{\mathrm Z}^2) 
             & = &  (280.9 \pm 6.3)\times 10^{-4}
        ~~~~(e^+e^-~\mathrm{data~only}) \nonumber \\
     \Delta\alpha_{\mathrm had}^{(5)}(M_{\mathrm Z}^2)
             & = &  (281.7 \pm 6.2)\times 10^{-4}
        ~~~~({\mathrm{combined}}~e^+e^-~\mathrm{and}~\tau~\mathrm{data})~.\nonumber
\eeqn
\par
F\/ig.~\ref{fig_results} shows a compilation of published results.
The inclusion of the new $\tau$ data yields a large
improvement in the precision of the \amuhad\ determination.
The dif\/ference in \amuhad\ between the exclusive \ee\ analyses
of~\cite{eidelman} and this work is mainly due to a disagreement in the
two-pion integral where we obtain signif\/icantly lower values. In 
addition, dif\/ferences in the handling of 
unmeasured modes generate inconsistent results. The results 
of Ref.~\cite{kinoshita} cannot easily be compared to the newer ones 
as the data set did not include the recent DM2 results. The dif\/ferences 
in the f\/inal errors of \amuhad\ in the exclusive \ee\ analysis of 
this work compared to Ref.~\cite{eidelman,worstell} is mainly caused by 
dif\/ferent techniques in the handling of the data and their errors.
The detailed study of the origin of correlations, their
propagation, as well as the rigourous use of isospin constraints
to bound unmeasured modes yield slightly smaller errors here.
As expected, the gain in the precision of \amuhad\ coming from $\tau$ data 
is signif\/icant whereas it is very small for \daqedhZ\ since the dominant 
contributions and uncertainties come from energies above the $\tau$ mass.

%
%
\subsection*{Higher Order Contributions}

In the famous paper of Kinoshita {\it et al}.~\cite{kinoshita}, higher order
contributions to the muon hadronic vacuum polarization graph, such as additional 
lepton or quark loops inserted in the diagram of Fig.~\ref{fig_amu}
and the so-called light-by-light scattering graph, have been evaluated.
The latter has been re-computed by dif\/ferent groups obtaining
$(-5.2\,\pm\,1.8)\times10^{-10}$~\cite{light1} and
$(-9.2\,\pm\,3.2)\times10^{-10}$~\cite{light2} with large
uncertainties compared to the designed experimental
accuracy of $\Delta a_\mu\simeq4.0\times10^{-10}$ of the forthcoming BNL 
experiment. We use the value of $(-6.2\,\pm\,4.0)\times10^{-10}$ in the 
following with an enlarged error to account for inconsistencies.

The calculation of the higher order ${\cal O}(\alpha/\pi)^3$ loop 
diagrams is accomplished and partly corrected in a recent 
work~\cite{krause2}, where second order kernel functions $K^{(2)}(s)$ 
are provided. These are used to calculate the corresponding 
contributions in the same spirit as the dominant lowest order graph 
by virtue of the dispersion integral~(\ref{eq_integral1}).
The numerical evaluation in Ref.~\cite{krause2} was performed on the basis
of the data sample used by Ref.~\cite{eidelman}. We repeat this exercise here
in order to check the consistency of the results. For the contribution
of diagrams with additional photon exchanges, {\it e.g.}, the fourth 
order muon vertex correction, we use the kernel labeled $K^{(2a)}(s)$ 
in~\cite{krause2} and obtain $a_\mu^{(2a)}\,=\,(-20.9$\pms$0.4)\times10^{-10}$.
The diagrams with an electron loop inserted in one of the photon lines
of Fig.~\ref{fig_amu} (kernel $K^{(2b)}(s)$ in Ref.~\cite{krause2}) contribute 
to $a_\mu^{(2b)}\,=\,(10.6$\pms$0.2)\times10^{-10}$, where the asymptotic
expansion, the analytical and numerical solutions provided in Ref.~\cite{krause2}
lead to very similar results. F\/inally, the insertion of two hadronic loops 
in the muon vertex correction graph (kernel $K^{(2c)}(s)$ in Ref.~\cite{krause2}) 
results in $a_\mu^{(2c)}\,=\,(0.27$\pms$0.01)\times10^{-10}$. The 
contributions $a_\mu^{(2a,b,c)}$ are found to be in agreement 
with Ref.~\cite{krause2}. All higher order results given here are 
computed from the \ee\ data set only.

The compilation of the hadronic higher order parts (including light-by-light 
scattering) yields \amuhad$[(\alpha/\pi)^3]\,=\,(-16.2 \,\pm\,4.0)\times10^{-10}$.

%
%
\subsection*{Results for \boldmath$a_\mu$ and \boldmath\aqedZ}

Collecting all contributions, we obtain for the anomalous magnetic moment 
of the muon
\beqn
      a_\mu & = & (11\,659\,164.5 \,\pm\, 15.6)\times10^{-10}
        ~~~~(e^+e^-~\mathrm{data~only}) \nonumber \\
      a_\mu & = & (11\,659\,170.6 \,\pm\, 10.2)\times10^{-10}
        ~~~~({\mathrm{combined}}~e^+e^-~\mathrm{and}~\tau~\mathrm{data})~,\nonumber
\eeqn
where the errors of the lowest order calculation \amuhad\ and 
$a_\mu^{(2a,b,c)}$ are added linearly.
\vs
The inverse of the f\/ine structure constant at $M_{\mathrm Z}^2$ is
found to be
\beqn
     \alpha^{-1}(M_{\mathrm Z}^2) & = & 128.889\,\pm\,0.087
        ~~~~(e^+e^-~\mathrm{data~only}) \nonumber \\
     \alpha^{-1}(M_{\mathrm Z}^2) & = & 128.878\,\pm\,0.085
        ~~~~({\mathrm{combined}}~e^+e^-~\mathrm{and}~\tau~\mathrm{data})~.\nonumber
\eeqn
\par
One may use the latter (combined) result for $\alpha(M_{\mathrm Z}^2)$ 
to improve the constraint on the mass of the standard model Higgs 
boson $M_{\mathrm{Higgs}}$ inferred from a global electroweak f\/it. 
This is done by utilizing current available electroweak 
data~\cite{Blondel,lepreport} and the ZFITTER electroweak 
library~\cite{leplib}. It requires $M_{\mathrm Z}$, $m_{\mathrm{top}}$, 
and $\alpha(M_{\rm Z}^2)$ as input parameters, which are allowed 
to vary within their experimental accuracies. The additional parameters 
$M_{\mathrm{Higgs}}$ and the strong coupling constant at $M_{\mathrm Z}^2$, 
$\alpha_s(M_{\mathrm Z}^2)$, are freely adjusted in the f\/it. We obtain 
$\alpha_s(M_{\mathrm Z}^2)\,=\,0.1201\,\pm\,0.0033$ which is in perfect 
agreement with the experimental value of 0.122\pms0.006~\cite{alphas1}
from the analyses of QCD observables in hadronic Z decays at LEP.
The f\/itted Higgs boson mass is $138^{+137}_{-76}$~GeV, compared 
to $149^{+148}_{-82}$~GeV when using the previous value of \aqedZ\ 
from Eq.~(\ref{eq_alpha}). An additional error of 50~GeV should be 
added to account for theoretical uncertainties~\cite{leplib}.
\begin{figure}[t]
\epsfxsize10cm
\centerline{\epsffile{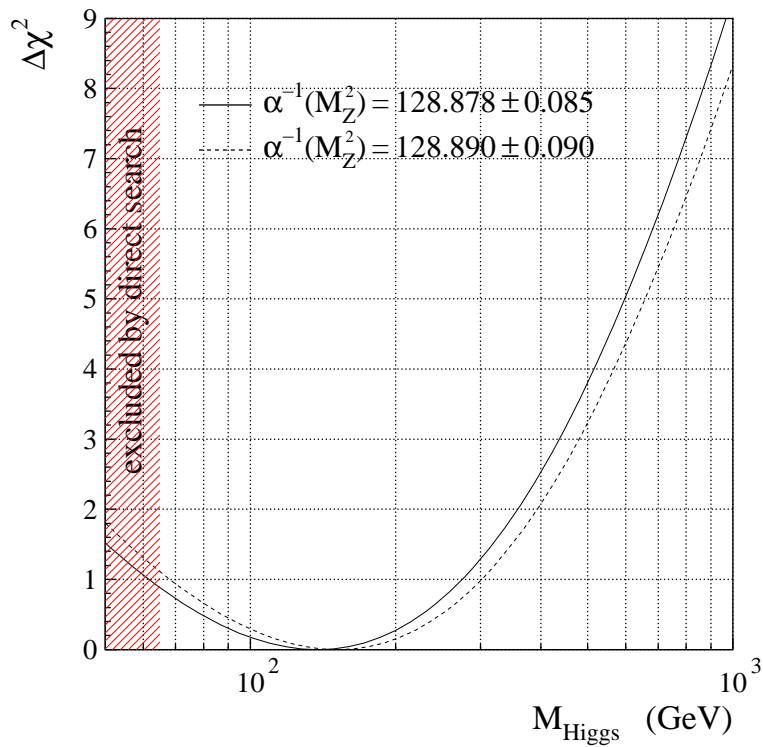}}
\caption[.]{\it Constraint fit results for the previous and the 
             new value of \aqedZ\ as a function of the Higgs mass.}
\label{fig_mhiggs}
\end{figure}

Fig.~\ref{fig_mhiggs} depicts the variation 
of $\chi^2$ as a function of the Higgs boson mass for the new and 
previously used values of \aqedZ\ (the latter taken from~\cite{eidelman}). 
We obtain an upper limit for $M_{\mathrm{Higgs}}$ of 516~GeV at 95\pc\ CL.

%
%
\section{Conclusions}

We have reevaluated the hadronic vacuum polarization contribution to 
$(g-2)$ of the muon and to the running of the QED f\/ine structure constant
$\alpha(s)$ at $s=M_Z^2$. We used new data from $\tau$ decays, recently
published by the ALEPH Collaboration, in addition to slightly enlarged \ee\
annihilation cross section data sets in order to improve the precision of 
the corresponding integrals. Our results are, to lowest order,
\amuhad\,=\,(701.1$\,\pm\,$9.3)$\times10^{-10}$ yielding 
$a_\mu\,=\,(11\,659\,170.6\,\pm\,10.2)\times10^{-10}$ and 
\daqedhZ\,=\,(281.7$\,\pm\,$6.2)$\times 10^{-4}$, propagating
$\alpha^{-1}(0)$ to $\alpha^{-1}(M_{\mathrm Z}^2)\,=\,(128.878\,\pm\,0.085)$.
The improvement coming from $\tau$ data is small in the \daqedhZ\ case 
which is dominated by high energy contributions. However, it causes a 
37\pc\ reduction in the error on \amuhad.
\vs
In the near future, new low energy \ee\ annihilation data are expected 
to be produced by the CMD-2 Collaboration~\cite{eidelmantau} 
at Novosibirsk. In addition, new results for the \tauto\pipiz\nut\ \sf\ 
with a precision comparable to the ALEPH data were recently presented by 
the CLEO Collaboration~\cite{cleo2}. Signif\/icant improvement is also
expected from energy scans at the future high-luminosity \ee\ collider 
DA$\Phi$NE in Frascati.

%
%
\section*{Acknowledgements}

It is a pleasure to thank A.~Blondel, H.~Burkhardt, S.~Eidelman, L.~Duflot, 
M.~Finkemeier, F.~Jegerlehner, B.~Krause, W.~Marciano, A.~Pich, B.~Pietrzyk 
and F.~Teubert for helpful comments and interesting discussions.
We are indebted to S.~Sen for carefully reading this manuscript.
%
%
\end{document}